\documentstyle[11pt,newpasp,twoside,epsf]{article}
\def\edcomment#1{\iffalse\marginpar{\raggedright\sl#1\/}\else\relax\fi}
\reversemarginpar

\begin{document}

\title{Dust disks around old Pre Main-Sequence stars: HST/NICMOS2
  scattered light images and modeling}
\author{Jean-Charles Augereau, Anne-Marie Lagrange, David Mouillet}
\affil{Laboratoire d'Astrophysique de l'Observatoire de Grenoble,
  Universit\'e J. Fourier, 414 rue de la piscine, BP 53,
  38041 Grenoble Cedex 9, France} \author{Fran\c{c}ois M\'enard}
\affil{CFH Telescope Corporation, PO Box 1597,
  Kamuela, HI 96743, USA}

\begin{abstract}
  We present recent near-infrared detections of circumstellar disks
  around the two old PMS Herbig stars HD\,141569 and HD\,100546
  obtained with the HST/NICMOS2 camera. They reveal extended
  structures larger than 350--400\,AU in radius. While the HD\,100546
  disk appears as a continuous disk down to 40\,AU, the HD\,141569
  environment seems more complex, splitted at least into two dust
  populations. As a convincing example, the full modeling of the disk
  surrounding HR\,4796, another old PMS star, is detailed and
  confronted with more recent observations.
\end{abstract}

\section{Introduction}
The presence of massive disks with large gas to dust ratios around
T\,Tauri stars is now well established (see a recent review by Dutrey
1999). Most recent near-IR and millimeter resolved images reveal
optically thick disks in Keplerian rotation around single or multiple
systems (e.g. GG Tau, Guilloteau et al.\,1999). The dust appears to be
the direct remnant of the primordial nebula and in a theoretical point
of view, Weidenschilling et al.\,(1993) have shown that planetesimals
may form during this stage.

It is also known that a large fraction of Main-Sequence (MS) stars are
surrounded by material heated by their central star (see a recent
review by Lagrange et al.\,2000). Among these Vega-like stars, the
$\beta$ Pictoris circumstellar disk has been for a long time the only
one imaged till recently.  Submillimetric observations resolved for
the first time dust emission around a few isolated MS stars (e.g.
Greaves et al.\,1998).  Vega-like stars bear optically thin disks with
small gas to dust ratios and continuously replenished in smallest
grains by collision and/or evaporation among larger bodies (Backman et
Paresce\,1993).  We have also direct (e.g.  55\,Cnc, Butler et
al.\,1997) or indirect ($\beta$ Pictoris: e.g. Beust \&\ 
Morbidelli\,\,2000) evidences for planets in these systems.

To better understand the full evolutionary scenario which lead to
planetary systems formation, one must answer the following questions:
what are the time-scales for dissipating disks? How does it depend on
the spectral type and multiplicity? What are the dynamical dominant
processes? How do gas and dust couple?  What is the origin and nature
of the dust?  And finally, how does it correlate with planet
formation? To do so, one must study circumstellar disks of
intermediate ages which represent the missing link between embedded
young disks and evolved ones. Full spectral energy distributions
(SEDs) give precious constraints on the chemical composition of the
grains (e.g. HD\,100546, Malfait et al.\,1998 with ISO) but only
qualitative informations on the dust distribution itself. Resolved
data are needed on these objects.

In section 2, we present HST/NICMOS2 images of two Herbig (Vega-like?)
stars: HD\,141569 and HD\,100546. We then discuss in section 3 how we
can infer valuable constraints on the dust composition and
distribution as well as on the dynamics of the system by combining all
available data (SED and multi-$\lambda$ images) through a consistent
model. The case of HR\,4796 will be detailed.
\section{HST/NICMOS2 observations of HD\,141569 and HD\,100546}
\label{obs}
\subsection{Targets and coronagraphic data}
HD\,141569 is a B9.5Ve star older than 10\,Myr and $99^{+9}_{-8}$\,pc
away ({\it Hipparcos} measurements, Van den Ancker et al.\,1998). This
star was selected due to its IRAS infrared (IR) excess and also to the
small intrinsic polarization (Yudin \&\ Evans\,1998) which are both
clues for suspecting the presence of an extended optically thin disk.
Moreover, the $^{12}$\,CO\,\,$J$=2-1 detection by Zuckerman et
al.\,(1995) ensures that gas is present at a level consistent with a
young MS star.

The Herbig B9V star HD\,100546, associated to the dark cloud DC
296.2-7.9, shows a strong IR excess peaked at about 25\,$\mu$m (IRAS)
due to a large amount of circumstellar material. The interpretation of
photometric, polarization and spectroscopic events (e.g. Van den
Ancker et al.\,1998; Yudin \&\ Evans\,1998; Grady et al.\,1997)
remains uncertain but are characteristic of Herbig Ae/Be stars.
HD\,100546 is $103^{+7}_{-6}$\,pc away and 10\,Myr old according to
Van den Ancker et al.\,(1998) and appears to be in a less evolved
state than HD\,141569.

We have performed, $\lambda$=1.6$\,\mu$m coronagraphic images of
HD\,141569 and HD\,100546 with the HST/NICMOS2 camera.  During the
{\it same} orbit (so as to avoid PSF variations), a close star free of
known circumstellar material so far and with similar spectral type and
magnitude was observed. The reduction procedure consists in
subtracting this reference star carefully scaled to the star of
interest.  The determination of the scaling factor is critical since a
small change can significantly modify the photometry or in the worst
case be responsible for the presence of not realistic features.  It is
found by azimuthally averaging the division between the star of
interest and the reference star images.
\subsection{Results for HD\,141569}
\begin{figure}
\plotfiddle{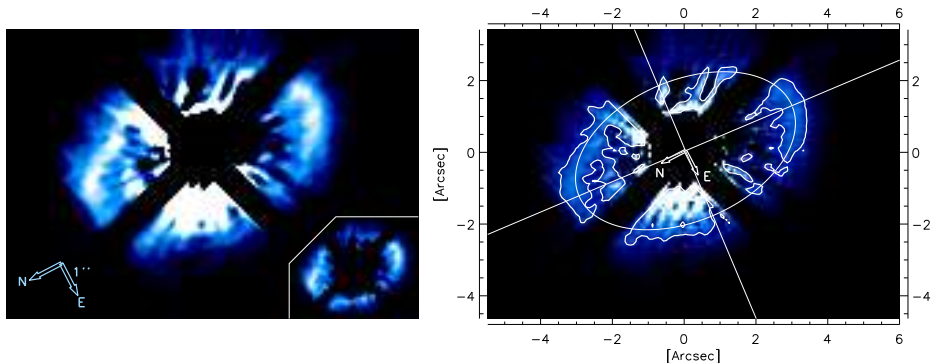}{110pt}{90}{32}{32}{48}{-25}
\caption{\small Left panel: scattered light image (1.6\,$\mu$m) of
  the HD\,141569 circumstellar disk in logarithmic scale (from
  Augereau et al.\,1999b). In the bottom right corner, we show the
  same disk where most of the unrealistic bright patterns close to the
  star have been removed so as to highlight the annular resolved
  structure. In the right panel, result of ellipse fitting.}
\end{figure}
The final reduced image of HD\,141569 shown in the left panel of
Figure 1 is interpreted as an inclined dust ring peaked at
325$\pm$10\,AU from the star which scatters a small fraction (a few
$10^{-3}$) of the stellar light. We derive from least-squares ellipse
fitting (Figure 1, right panel) a position angle (PA) of
355.4$^{\mathrm o}\pm$1$^{\mathrm o}$ and a disk inclination from
edge-on of 37.5$^{\mathrm o}\pm$4.5$^{\mathrm o}$ assuming that the
disk is axisymmetrical with respect to the star. The radial surface
brightness indicates that the outer edge of the ring steeply decreases
with the distance from the star following a $r^{-6.87\pm0.14}$ radial
power law between 360\,AU and 420\,AU.

The presence of an inclined extended disk-like structure around
HD\,141569 has been confirmed independently by Weinberger at
al.\,(1999) through HST/ NICMOS2 1.1\,$\mu$m images. The
interpretation of the inner part of disk (below $\sim$ 240\,AU along
the major axis) is controversial. They found that most of the
scattered light comes from a region rather peaked at $\sim$ 200\,AU
from the star.  Scattering properties cannot account for the
difference between our 1.6\,$\mu$m and 1.1\,$\mu$m images. At
1.6\,$\mu$m, bright areas which appear inside the inclined annulus are
interpreted as unrealistic residues to PSF subtraction. This
interpretation is supported by the fact that the ratio of HD\,141569
(before subtraction) to the reference star clearly evidences a lack of
significant detected circumstellar material inside the main resolved
shape (see Fig. 2 of Augereau et al.\,1999b).
\subsection{Results for HD\,100546}
\begin{figure}
\plotfiddle{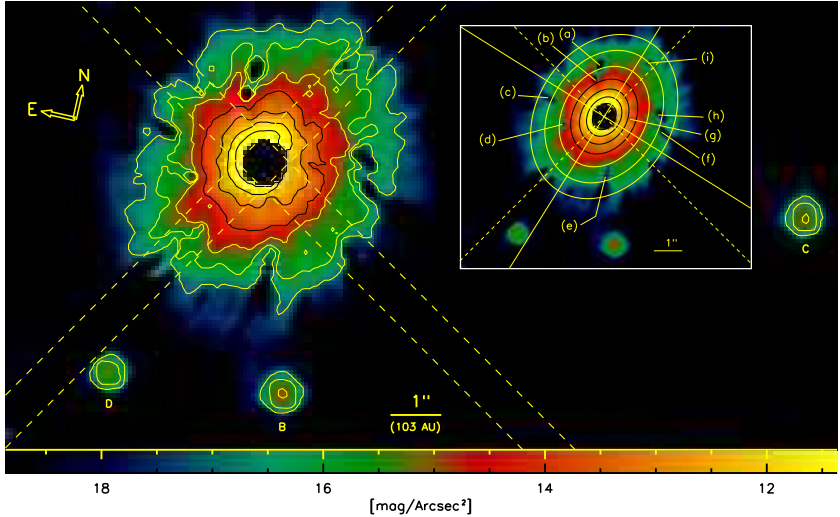}{175pt}{-90}{50}{50}{-210}{236}
\caption{\small Scattered light image (1.6\,$\mu$m) of
  the HD\,100546 circumstellar disk (from Augereau et al.\,2000). The lowest
  contour level is 17\,mag/arcsec$^2$. The isophotes spacing is
  1\,mag/arcsec$^2$ in increasing order. The small panel in the upper
  right corner highlights both the fix patterns (labeled $(a)$ to
  $(i$)) which blurres the image and the result of isophotes ellipse
  fitting. Plain lines: major and minor axis of the disk; dashed
  lines: position of spider arms.}
\end{figure}
The reduced image of HD\,100546 shown in Figure 2 also reveals a large
elliptical circumstellar structure centered on and bright close to the
star. This observation seems consistent with preliminary results from
Pantin et al.\,(2000) obtained with the adaptive optics system ADONIS
(ESO). Contrary to HD\,141569, the HD\,100546 disk does not shape a
ring-like morphology but rather exhibits a continuous surface
brightness distribution from the very close edge of the coronagraphic
mask ($\sim$40\,AU) up to $\sim$350\,AU. Isophotes ellipse-fitting
constrains the PA of the disk to be 161$^{\mathrm o}\pm$5$^{\mathrm
  o}$ and provides an upper limit of 51$^{\mathrm o}\pm$3$^{\mathrm
  o}$ for the inclination of the disk with respect to the line of
sight. The measured flux density of 84$\pm$8\,mJy leads to a lower
limit of $10^{-2}\pm 15$\% to the scattered to photospheric flux,
assuming that 55\% to 60\% of the total flux (star+disk) at
1.6$\,\mu$m is due to thermal emission of very hot unresolved grains.

Basic radial power-laws fit very well the measured surface brightness
profile with indexes -2.92$\pm$0.04 in the radial range
40\,AU--250\,AU and -5.5$\pm$0.2 further 270\,AU. This implies a
surface density roughtly proportional to $r^{-1}$ which coincides with
results for less massive young stars (Dutrey et al.\,1999 and
references therein).  Subtracting a synthetic disk, supposed to
exhibit an axisymmetrical surface brightness as described above and
inclined at 51$^{\mathrm o}$ towards PA=161$^{\mathrm o}$, to the
observed image reveals a NE-SW brightness asymmetry: more precisely,
an excess of flux up to 1.1\,mag/arcsec$^2$ at 0.5--0.6$\arcsec$
collimated in a direction which coincides well with the precise minor
axis of the disk.  Anisotropic scattering dust grains properties can
explain this effect (e.g. Fig. 12 of Augereau et
al.\,1999a).
\section{Modeling of dust disks around old PMS stars}
\subsection{Short overview of model assumptions}
We developed an optically thin disk model able to reproduce, in a
consistent way, scattered and thermal observations as well as SEDs.
The dust distribution is described by a smooth combination of radial
power-laws. The vertical distribution is assumed to follow an
exponential shape with height depending on the distance from the star
as a radial power-law. A key step in the modeling are the grain
optical properties. We assume that grains are porous aggregates made
of a silicate core coated by an organic refractory mantle
(Greenberg\,1986). Vacuum is assumed to fill the holes due to porosity
which can be partially taken up by H$_2$O ice if the grain temperature
falls below 110-120\,K. We adopt two types of grains: amorphous
"ISM-like grains" with porosity of about 50\% and crystalline
"comet-like grains" with large porosities ($\sim$ 95\%).  Dust optical
properties are computed using Mie theory.  They depend on the complex
index of refraction of the mixture and on the grain size $a$. We adopt
the Maxwell-Garnett effective medium theory to compute these indexes
and assume a collisional grain size distribution following a
$a^{-3.5}$ power-law between $a_{min}$ and $a_{max}$. A more detailed
description of model assumptions can be found in Augereau et
al.\,(1999a).
\subsection{An example: the HR\,4796 disk}
HR\,4796 is a wide binary A0V star showing an infrared excess twice
that of $\beta$ Pictoris and 5 to 15 times younger ($8\pm 2$\,Myr
according to Stauffer et al.\,1995). HR\,4796 thus traces an
evolutionary stage prior to that of $\beta$ Pictoris. We used the
model described above to reproduce all available observations of the
HR\,4796 circumstellar system before 1999 (Augereau et al.\,1999a). We
find that two dust populations are required to account for the full
SED compatible with thermal and scattered light resolved images.

The first population shapes a sharp and cold ring peaked at 70\,AU
from the star and made of "ISM-like grains". It is responsible for the
25\,$\mu$m--1\,mm dust continuum emission (see Fig. 11 of Augereau
et al.\,1999a). This implies grains larger than $a_{min}=10\,\mu$m
which exactly corresponds to the blow-out size limit for grains
produced by larger bodies on circular orbits. Also, such large bodies
must be present either to reproduce millimeter measurements and to
resupply the ring in smallest grains. Indeed, the 10\,$\mu$m grains
have very short time-scales under collisions ($\sim$ 1000\,yr) compared
to the star age.

The second hot dust population is poorly constrained but would rather
be composed of large crystalline grains at 9--10\,AU from the star.
This population is necessary to fit the 10\,$\mu$m measurements which
can not be reproduced with a single ring at 70\,AU as sharp as
required by scattered light images (Schneider et al.\,1999).
Interestingly, Telesco et al.\,(2000) latter marginally resolved the
HR\,4796 disk in N band ($\lambda=10.8\,\mu$m, $\Delta
\lambda=5\,\mu$m) and show that a significant fraction of this
emission comes from the outer ring.  To check the consistency of
proposed model with these observations, we convolve with the observed
PSF from Telesco et al.\,(2000) a simulated disk at 10.8\,$\mu$m. We
also simulate a second disk at 12.8\,$\mu$m since the N band is wide
and since the disk emission across this band varies considerably
(Fig. 11 of Augereau et al.\,1999a). Figure 3 shows that the outer
ring detection is predicted by the model at a level fully consistent
with the observations given the spectral width of the N band. This
implies in particular that the relative fractions of flux coming from
the inner and outer disks are correctly predicted.
\section{Concluding remarks}
Further modeling, and especially dynamical modeling, is necessary to
improve our knowledge of disks at different stages. In the case of
HR\,4796, Wyatt et al.\,(1999), Kenyon et al.\,(2000) and Klahr et
al.\,(2000) have proposed models either to explain the ring-shaped
structure nor brightness asymmetries. Further studies of HD\,141569
and HD\,100546 will allow to evidence which dynamical processes are
generic or particular to such old PMS stars. But actually, very few
observations of these circumstellar environments are available.  As a
new example, SED fitting of the HD\,141569 disk using the radial dust
distribution derived from the 1.6\,$\mu$m images has predicted the
presence of an inner dust population (Augereau et al.\,1999b),
recently confirmed by Fisher et al.\,(2000).
\begin{figure}
\plotfiddle{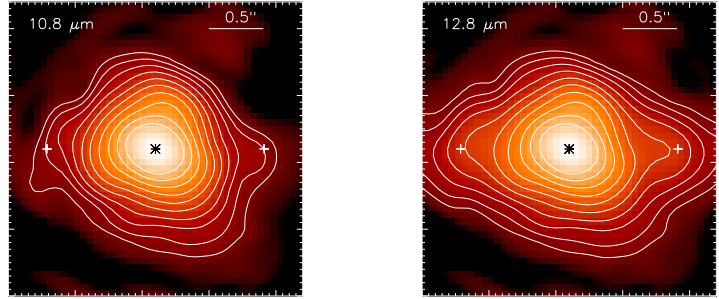}{90pt}{90}{55}{55}{250}{-58}
\caption{\small Simulated thermal images of the HR\,4796 disk convolved
  with the N band observed PSF from Telesco et al.\,(2000). To allow
  direct comparison, the contours levels are strictly the same as those
  of Figure 1a from these last authors. The white crosses indicate the
  position of outer ring.}
\end{figure}
\acknowledgments We thank Mark Wyatt for suggesting to check the
consistency of the HR\,4796 model with Telesco et al.\,(2000)'s
10\,$\mu$m observations and Charles Telesco for kindly providing the
corresponding observed PSF.
\end{document}